\documentclass[fleqn]{llncs}
\usepackage{version}
\usepackage{os}

\title{Searching Publications on Operating Systems}
\author{C.A. Middelburg}
\institute{Informatics Institute, Faculty of Science,
           University of Amsterdam, \\
           Science Park~107, 1098~XG Amsterdam, the Netherlands \\
           \email{C.A.Middelburg@uva.nl}}

\begin{document}
\maketitle

\begin{abstract}
This note concerns a search for publications in which one can find
statements that explain the concept of an operating system, reasons for
introducing operating systems, a formalization of the concept of an
operating system or theory about operating systems based on such a
formalization.
It reports on the way in which the search has been carried out and the
outcome of the search.
The outcome includes not only what the search was meant for, but also
some added bonuses.
\end{abstract}

% operating system, bibliography, journal collection, search engine.
% D.4.m

\section{Introduction}
\label{sect-intro}

The study of various issues raised in theoretical work on operating
systems and practical work in which operating systems are involved, such
as work in the area of digital forensics, calls for an abstract model of
an operating system that can be considered an adequate formalization of
the concept of an operating system.
Moreover, a certain body of theory about operating systems based on the
model in question would be helpful in studying the issues concerned.
To check whether a more or less appropriate model, as well as theory
based on it, already exists, I carried out a search for witnessing
publications.

I ran across some specifics of the search by which an ad hoc approach to
it would give little confidence in its outcome.
This brought me to devise a more or less systematic way to carry out the
search.
In addition to what was looked for, the search yielded added bonuses.
This note reports on the way in which the search has been carried out
and the outcome of the search, including the added bonuses.

\section{The Approach to Carry out the Search}
\label{sect-search}

For the search, Google Scholar and the search engines of relevant
journal collections and bibliographies are available.
Examples of relevant journal collections are ACM Digital Library,
SpringerLink, ScienceDirect, and IEEE Xplore.
Examples of relevant bibliographies are ACM Guide, DBLP Computer Science
Bibliogra\-phy and the Collection of Computer Science Bibliographies.
The search engines of relevant journal collections and bibliographies
have the option to search for publications in which given terms occur in
their title, abstract or keywords, but none cover all relevant
literature.
Google Scholar does not have this option, i.e.\ only full text search is
possible, but it covers all literature covered by the search engines of
relevant journal collections.

In a preparatory phase of the search, we found the following specifics:
\begin{itemize}
\item
the term ``operating system'' occurs in extremely many publications
whose subject is not operating systems;
\item
in the early days of operating systems, the term ``operating system'' is
not used in publications on operating systems;
\item
till 1970, a large part of the publications on operating systems appear
in the Communications of the ACM;
\item
from 1969, a large part of the publications on operating systems appear
in ACM SIGOPS Operating Systems Review and the ACM proceedings series
SOSP. % Symposium on Operating Systems Principles
\end{itemize}

By the first two specifics, an ad hoc approach to the search would give
little confidence in its outcome.
The first specific calls for a search for publications in which the term
``operating system'' occurs in their title, abstract or keywords,
but without special measures this will result in poor coverage of the
relevant literature.
The second specific complicates a search because not all terms used
instead of ``operating system'' need to be known beforehand, and some
terms, such as ``supervisor'' and ``monitor'' occur often in different
meanings.
I took the last two specifics as means to deal with the problems that
arise from the first two specifics.
This led to the following more or less systematic way to carry out the
search:
\begin{itemize}
\item
first, search for relevant publications by having a look at all
publications in all issues of Communications of the ACM till 1970,
all issues of ACM SIGOPS Operating Systems Review, and all SOSP
proceedings;
\item
next, search for relevant publications by means of the search engines of
relevant journal collections and bibliographies;
\item
then, search for relevant publications by means of Google Scholar;
\item
check immediately after finding a relevant publication whether it
references directly or indirectly additional relevant publications;
\item
stop searching by means of Google Scholar when 500 consecutive hits
yields no new relevant publication.
\end{itemize}

The number of publications to be looked at in the first step is about
7,000 (additional publications found by following references not
counted).
The sum of the numbers of publications on operating systems found by the
different search engines of relevant journal collections and
bibliographies is about 24,000, but there are many duplicates.
I first searched with the search engine of DBLP Computer Science
Bibliography, which actually turned out to yield most of the relevant
publications found by means of the search engines of relevant journal
collections and bibliographies.
The number of publications found by a full text search for the term
``operating system'' with Google Scholar is about 500,000.
I looked only at the first 1,000 hits, because Google Scholar yielded
very few new relevant publications.

\section{A Summary of the Main Outcome of the Search}
\label{sect-summary-outcome}

I searched for publications in which one can find statements that
explain the concept of an operating system, reasons for introducing
operating systems, a formalization of the concept of an operating system
or theory about operating systems based on such a formalization.
It turned out that the number of such publications is very small.

In~\cite{CLMS59a}, Codd and others give motivation for, requirements
for, and functions of a multiprogramming operating system.
This can be taken for a preparation to the formulation of the scheduling
problem in multiprogramming operating systems in~\cite{Cod60a} and the
description of a scheduling algorithm for a multiprogramming operating
system in~\cite{Cod60b}.
Probably these three articles belong to the first important articles on
operating systems.
It is often stated that Strachey's article on multiprogramming operating
systems~\cite{Str59a} is the first important article on multiprogramming
operating systems, but it appears that a more moderate statement is more
appropriate.%
\footnote
{Strachey's paper can only be obtained by ordering a hard copy at the
 National Archive of the United Kingdom.}

Apart from the attempt of Codd and others in~\cite{CLMS59a}, few serious
attempts have been made to explain the concept of an operating system.
Dennis and Van Horn make a serious attempt in~\cite{DH66a} and Denning
makes another serious attempt in~\cite{Den71a}, but most other attempts
cannot be called serious.
Examples of non-serious attempts are one-liners like ``an operating
system is an extended machine and a resource manager'' and enumerations
of the usual terms for the basic constituents of an operating system.

Apart from the reasons given by Codd and others in~\cite{CLMS59a},
reasons for introducing operating systems are seldom given.
Cloot gives good reasons in~\cite{Clo65a}, an article whose sole aim is
to explain why the need for operating systems has arisen, but usually
the reasons are not more advanced than ``it is useful to have an
operating system available''.

In~\cite{YLSL99a}, Yates and others give an abstract model of an
operating system, using input/output automata, which looks to be a model
that can be considered an adequate formalization of the concept of an
operating system.%
\footnote
{The paper of Yates and others actually gives two models.
 The abstract model is the model that is called the user level model in
 the paper.
}
Apart from this, publications in which abstract models of an operating
system are given that can be considered an adequate formalization of the
concept of an operating system are virtually absent.
In~\cite{Hei64a}, Heistand gives an abstract model of a part a specific
time-sharing operating system, using finite-state automata.
It is questionable whether this model can be considered an adequate
formalization of the concept of an operating system.
In~\cite{DK80a}, Degtyarev and Kalinichenko give a very abstract model
of an operating system.
Because it abstracts also from essentials of an operating system, this
model cannot be considered an adequate formalization of the concept of
an operating system.
Neither the queueing-theoretical models of process scheduling in
time-sharing operating systems, such as the models of Kleinrock and
Coffman~\cite{Kle67a,CK68a}, are the models we have in mind.

In~\cite{McK62a}, McKenney mentions that a conceptual model of an
operating system based upon analysis of operating systems and a
literature study is given in his Ph.D.\ thesis.
Little more is known about this model: the thesis is not cited in other
publications and whether it can still be obtained is doubtful.
In any case, it is unlikely that the model of McKenney is not superseded
by the model of Yates and others mentioned in the previous paragraph.

Publications on theory about operating systems themselves are totally
absent.
In publications on operating systems that are of a theoretical nature,
one finds only theory about internals of operating systems such as
process scheduling and resource allocation.
For early articles presenting such theory, see e.g.~\cite{Kle67a,CK68a}
and~\cite{DGP67a,Den68a}, respectively.

From this outcome of the search, we conclude that there exists only one
abstract model of an operating system that can be considered a more or
less appropriate formalization of the concept of an operating system,
viz.\ the abstract model given by Yates and others in~\cite{YLSL99a},
and that there does not exist theory
based on that model.
Moreover, we conclude that the operating systems community pays little
attention to clarifying adequately what is an operating system and
giving motives for introducing operating systems.
All this raises the question what most publications on operating systems
are about.
It turns out that they mainly concern the following:
\begin{itemize}
\item
principles of operating system design;
\item
theory and techniques related to internals of operating systems such as
process scheduling and resource allocation;
\item
issues concerning operating systems for multi-processor computers and
operating systems for networks of distributed computers;
\item
operating system support for security, privacy, fault-tolerance,
multi-media applications, etc.;
\item
designs of, analyses of, and experiences with specific operating
systems.
\end{itemize}
It is striking that most of these publications give little insight in
the concept of an operating system.
My findings of the search agree with the finding of the study of courses
and textbooks presented in~\cite{CS00a}.

\section{The Added Bonuses of the Search}
\label{sect-added-bonuses}

I found many statements that try to explain the concept of an operating
system in ways that lack generality, abstractness or preciseness.
However, I could easily distill the essence from them.
This led to the following explanation of the concept of an operating
system.

An operating system is a system that provides a convenient execution
environment for programs that allows for multiple programs with shared
resources to be executed concurrently.
An operating system is responsible for:
\begin{enumerate}
\item
loading programs and starting their execution;
\item
scheduling the programs in execution;
\item
allocating resources to the programs in execution;
\item
preventing interference between the programs in execution;
\item
controlling the use of main memory by the programs in execution;
\item
storing and retrieving data organized into files and directories on
secondary storage devices;
\item
receiving data from input devices and sending data to output devices;
\item
communicating data over computer networks;
\item
controlling peripheral devices.
\end{enumerate}

It is customary to distinguish the following basic constituents in an
operating system:
\begin{itemize}
\item
process management, responsible for 1, 2, 3 and 4;
\item
memory management, responsible for 5;
\item
file management, responsible for 6;
\item
input/output management, responsible for 7;
\item
network management, responsible for 8;
\item
device management, responsible for 9.
\end{itemize}
Process management and a part of memory management are needed to provide
an execution environment for programs that allows for multiple programs
with shared resources to be executed concurrently.
Device management, network management, input/output management, file
management, and a part of memory management are needed to provide a
\emph{convenient} execution environment, because they hide interrupts,
networking protocols, device-dependent input, output and storage,
physical memory size, etc.

Operating systems can be classified as:
\begin{itemize}
\item
single-user or multi-user;
\item
non-interactive or interactive;
\item
single-tasking, non-preemptive multi-tasking or preemptive
multi-tasking.
\end{itemize}
Actually, the explanation given above is an explanation of the concept
of an multi-tasking operating system.
Single-tasking operating systems are border cases of operating systems:
the maximal number of programs that can be executed concurrently is only
one.
Clearly, a multi-tasking operating system is a more general concept than
a single-tasking operating system.

Batch operating systems are multi-user, non-interactive,
single-tasking operating systems.
Multiprogramming operating systems are multi-user, non-inter\-active,
(non-preemptive or preemptive) multi-tasking operating systems.
Time-sharing operating systems are multi-user, interactive, preemptive
multi-tasking operating systems.
Ryckman mentions in~\cite{Ryc83a} that the Input/Output System for the
IBM~704 computer, which was developed by General Motors and North
American Aviation and became operational in 1956, is the first batch
operating system.
The Atlas Supervisor of Kilburn and others~\cite{KPH61a}, which was
developed over the period 1957--1961, is generally considered the first
multiprogramming operating system.
The experimental time-sharing operating system CTSS of Corbat{\'{o}} and
others~\cite{CMD62a}, which was developed over the period 1961--1963, is
generally considered the first time-sharing operating system.

I accidentally found additional interesting facts about the history of
operating systems which seem to be overlooked in all historical
overviews.

We already mentioned that Strachey's article on multiprogramming
operating systems~\cite{Str59a} is often mentioned as the first
important article on multiprogramming operating systems, but that there
exists an article by Codd and others, namely~\cite{CLMS59a}, which
probably belongs to the first important articles on multiprogramming
operating systems as well.
In the article of Codd and others is referred to~\cite{Bro57a}, which is
an article on a program interruption system published in 1957.
I found that the latter article already mentions multiprogramming
without explanation.
Therefore, I decided to search for earlier publications in which the
term ``multiprogramming'' occur.
I found that multiprogramming is mentioned as a programming technique
for sorting by merging in a paper by Friend on sorting~\cite{Fri56a}
published in 1956.
Moreover, multiprogramming is shortly explained by means of an example
in a paper by Rochester~\cite{Roc55a} published in 1955.
To my knowledge, these papers are never mentioned in historical remarks
made in the literature in question.

The idea of a time-sharing operating system is usually attributed to
McCarthy, who proposed to develop a time-sharing operating system at MIT
in 1959.
His recollections about this matter can be found in~\cite{LML92a}.
The term ``time-sharing'' has caused confusion about the origin of the
idea of a time-sharing operating system, because this term is used
before 1959 by Everett and others in~\cite{EZB57a} and Bemer
in~\cite{Bem57a} for preemptive multi-tasking in a non-interactive
setting.
In~\cite{Str59a}, Strachey uses the term ``time-sharing'' for
``multi-tasking'' as well.
On that account, Corbat{\'{o}} and others~\cite{CMD62a} attributes the
idea of a time-sharing operating system incorrectly to Strachey.

\section{Concluding remarks}
\label{sect-concl}

To check whether an abstract model of an operating system that can be
considered a more or less appropriate formalization of the concept of an
operating system and a certain body of theory about operating systems
based on it already exist, I have carried out a search for witnessing
publications.
To be able to judge the suitability of existing models, I have also
looked for publications that contain statements explaining the concept
of an operating system and publications that give reasons for
introducing operating systems.

From the outcome of the search, I conclude that there exists only one
more or less appropriate model, viz.\ the abstract model given by Yates
and others in~\cite{YLSL99a}, and that there does not exist theory based
on that model.
In addition, I conclude that the operating systems community pays little
attention to clarifying adequately what is an operating system and
giving motives for introducing operating systems.
However, I found many statements that try to explain the concept of an
operating system in ways that lack generality, abstractness or
preciseness.
By distilling the essence from those statements, I have obtained an
explanation of the concept of an operating system that could serve as
the starting-point of the development of an abstract model of an
operating system.
It happens that the above-mentioned operating system model of Yates and
others agrees with this explanation.

\bibliographystyle{spmpsci}
\bibliography{OS}

% \par \vfill \par \noindent DRAFT of \today

\end{document}